\documentstyle[aps,epsf,epsfig,prl,psfig]{revtex}
\newcommand{\beq}{\begin{equation}}
\newcommand{\eeq}{\end{equation}}
\newcommand{\beqas}{\begin{eqnarray*}}
\newcommand{\eeqas}{\end{eqnarray*}}
\newcommand{\beqar}{\begin{eqnarray}}
\newcommand{\eeqar}{\end{eqnarray}}

\begin{document}
 \newcommand \be {\begin{equation}}
\newcommand \ee {\end{equation}}

\twocolumn[\hsize\textwidth\columnwidth\hsize\csname
@twocolumnfalse\endcsname
\title{Rational Decisions, Random Matrices and Spin Glasses}
\author{Stefano Galluccio$^\dagger$, Jean-Philippe Bouchaud$^{\dagger,*}$ and Marc Potters$^\dagger$}
\address{$^\dagger$ Science \& Finance, 109-111 rue Victor Hugo, 92523 Levallois Cedex, FRANCE}
\address{$^*$ Service de Physique de l'\'Etat Condens\'e,
 Centre d'\'etudes de Saclay, \\ Orme des Merisiers, 
91191 Gif-sur-Yvette C\'edex, FRANCE\\}
\date{\today}
\maketitle
\widetext

\begin{abstract}
We consider the problem of rational decision making in the presence of 
nonlinear constraints. By using tools borrowed from spin glass and 
random matrix theory, we focus on the portfolio optimisation problem.
We show that the number of ``optimal'' solutions
is generically exponentially large: rationality is thus {\it de facto} of 
limited use. In addition, this problem is related to spin glasses with 
L\'evy-like (long-ranged) couplings, for which we show that the ground state 
is {\it not} exponentially degenerate.

\end{abstract}
]
\narrowtext

It is often hoped that science can help in choosing the right decision. A
rational decision is usually thought of as the solution which optimizes a
certain
utility (or cost) function, which is supposed to quantify the performance of
a given strategy. A simple example is that of {\it portfolio
optimization}: given a set of financial assets, characterized by their
average return and their risk, what is the optimal weight of each asset,
such that the overall portfolio provides the best return for a fixed level
of risk, or conversely, the smallest risk for a given overall return?
It is frequently the case that the optimal solution is unique, so that
`rational' operators should choose this particular one, unless they are
inefficient. In the case of portfolio selection, the only free parameter is
the proportion of the ``riskless asset''; all portfolios should thus look very
much alike. This is at the heart of the so called {\it Capital Asset Pricing
Model}, (CAPM), one of the cornerstones of modern theoretical finance
\cite{CAPM}. 

In this Letter we will show that, surprisingly, there are cases where the
number of
``rational'' decisions is exponentially large (for example in the number of
assets) -- much like the number of ground states in spin glasses or other
disordered systems \cite{MPV}. This means that there is an irreducible 
component of 
randomness in the final decision: the degeneracy between this very large
number of possibilities can only be lifted by small, `irrational', effects.

More precisely, the average return $R_P$ of a portfolio $P$ of
 $N$ assets, is defined as $R_P=\sum_{i=1}^N p_i R_i$,
where $p_i$ 
$(i=1,\cdots,N)$
is the amount of total capital invested in the
 asset $i$, and $\{R_i\}$ are the expected returns of the
individual assets. 
Similarly, the risk on a portfolio can be associated to the total
variance
$\sigma_P^2=\sum_{i=1}^N p_i^2\sigma_i^2+\sum_{i\ne j}^N p_ip_j \sigma_{ij}$,
where $\sigma_i^2$ are the single asset variances and $\sigma_{ij}$ are
the inter-asset covariances. Alternatively,  one can define 
$\sigma_P^2=\sum_{i,j=1}^N p_iC_{ij}p_j$ 
where $C_{ij}$ is the covariance matrix.  

Generally speaking, the goal of each rational investor consists in maximizing
the portfolio return and/or in minimizing its variance (or risk). This 
optimization procedure, or {\it portfolio selection}, as  
investigated in its simplest version by Markowitz \cite{markovitz},
results in a set of all  possible ``optimal'' couples
$\{R_P,\sigma_P^2\}$
which all lie on a curve
called the {\it efficient frontier} ({\sc ef}). 
For a given degree of risk, there is thus a unique
selection of assets 
$\{ p_i^*\}_{i=1}^N$ which maximizes the expected return.
On the opposite, we will show that in some cases 
this simple scenario does not hold: a very large
number of compositions $\{ p_i\}_{i=1}^N$ can be ``optimal'', in the sense
defined above.
Furthermore, these optimal compositions can be very different from another.

In order to illustrate this rather general scenario with an explicit example,
 we shall investigate
the problem of portfolio selection in the case one can buy, but also to short sell stocks, currencies, 
commodities and other
kinds of financial assets. This is the case of `futures' markets or margin 
accounts. The only
requirement is to leave a certain deposit (margin) proportional to the
value of the underlying asset \cite{Hull}. This means that the overall
position on these
markets is limited by a constraint of the form:
\be
\sum_{i=1}^N \gamma W_i |p_i| = {\cal W},
\ee
where $W_i$ is the price of asset $i$, $\cal W$ the total wealth of the
operator, and $\gamma$ the fraction defining the margin requirement. Here,
$p_i$
is the number of contracts on asset $i$ which are bought ($p_i > 0$) or sold
($p_i < 0$).  It is worth stressing that it is the nonlinear form
of the above constraint that makes the problem interesting, as we shall see
below. At this point we simply note that the optimal
portfolio $p_i^*$, which -- say -- minimizes the risk for a given return, can be
obtained using two Lagrange multipliers ($\nu,\mu$):
\be
\frac {\partial}{\partial p_i} \left[\frac{1}{2} \sum_{j,k=1}^N p_j C_{jk} p_k -
\nu
\sum_{j=1}^N p_j R_j - \mu \sum_{j=1}^N W_j |p_j| \right] =
0.
\ee
Without loss of generality, one can always set $W_i \equiv 1$. Defining
$S_i$ to be the sign of $p_i^*$, one gets:
\be
p_i^* = {\nu} \sum_{j=1}^N C^{-1}_{ij} R_j + \mu \sum_{j=1}^N C^{-1}_{ij} S_j,
\label{sol1} \ee
where {\bf C}$^{-1}$ is the matrix inverse of {\bf C}. Taking the sign of this
last equation leads to an equation for the $S_i$'s which is identical to those
defining the locally stable configurations in a spin-glass \cite{MPV}:
\be
S_i = \mbox{sign} \; \left [H_i + \sum_{j=1}^N J_{ij}
S_j\right],\label{tap}
\ee
where $H_i \equiv \nu \sum_j^N C^{-1}_{ij} R_j$ and
$J_{ij} \equiv \mu  C_{ij}^{-1}$ are the analogue of the ``local field'' 
and of the spin interaction matrix, respectively. 
Once the $S_i$'s
are known, the $p_i^*$ are determined by Eq. (\ref{sol1}); $\nu$ and
$\mu$ are fixed, as usual, so as to satisfy the constraints.  Let us consider,
for simplicity, the case where one is interested in the minimum risk
portfolio, which corresponds to $\nu=0$. [By analogy with spin glasses, the
following 
results are not
expected to change qualitatively if $\nu \neq 0$ \cite{Dean}].  In this case,
one sees that only $|\mu|$ is fixed by the constraint. Furthermore, the
minimum risk is given by ${\cal R}=\mu^2 \sum_{j,k}^N S_j C^{-1}_{jk} S_k$,
which
is the analogue of the energy of the spin glass. It is now well known that if {\bf J}
 is a random matrix, the so called
``{\sc tap}'' equation (\ref{tap}), defining the metastable states of a spin glass
at zero temperature, generically has an exponentially large (in $N$) number of
solutions \cite{MPV}. Note that, as stated above, 
the multiplicity of solutions  is a direct consequence of the nonlinear
constraint (1).

In what respect can the correlation matrix {\bf C} be
considered as random? Here we take a step in complete analogy with the 
original Wigner and Dyson's idea of replacing the Hamiltonian of a
deterministic but complex system by a random matrix \cite{WigDys}. More
precisely, 
they proposed to study the properties of one generic member of a statistical
ensemble 
\cite{Mehta} which shares the symmetries of the original Hamiltonian. 
Similarly, in the present situation, we would like to see {\bf C}$^{-1}$ as 
a
random matrix whose elements are distributed according to a given ensemble,
compatible with some general properties. 
In our case, for instance, we must select {\bf C}$^{-1}$ from an  ensemble of
positive definite matrices, as all the eigenvalues of the covariance matrix
are $>0$. The choice of a suitable statistical ensemble is guided by the following
observation. The (daily) fluctuations of the asset $i$, $\delta W_i$, can be
decomposed 
as:
\be
\delta W_i = \sum_{\alpha=1}^K M_{i\alpha} \delta E_\alpha + \delta W_{i0},
\ee
where the $E_{\alpha}$ are $K$ independent factors which affects the assets
differently, and $\delta
W_{i0}$ is the part of the fluctuation which is specific to asset $i$ (and
thus independent of the $E_{\alpha}$). The $E_{\alpha}$
contain all information on how the stochastic evolution of a given asset
is correlated to the others. The matrix $C_{ij}$ can thus be
written as {\bf C} $=$ {\bf M\, M}$^T$ + {\bf D}, where 
{\bf D} is a positive diagonal
matrix, and {\bf M} a $N \times K$ rectangular matrix. The above
representation 
ensures that all eigenvalues of {\bf C} are positive.
In order to simplify the problem, we shall make in the following the
assumption that 
{\bf D} is proportional to the identity matrix and that the coefficients
$M_{i\alpha}$ are completely random. For $K=N$, {\bf C} is a member
of the so-called {\it Exponential Orthogonal Ensemble} ({\sc eoe}) \cite{bronk}, which is
a maximum entropy 
(least information) ensemble. For any $N\times K$ matrix {\bf M}, (with $N \ge
K$), the 
 density of eigenvalues $\rho_C(\lambda)$ of ${\bf C}$ is exactly known in the
limit
 $N \to \infty$, $K \to \infty$ and $Q=K/N$ fixed \cite{Sengupta}. In the
limit $Q=1$ 
(which we consider from now on) the normalized eigenvalue density 
 of the (square) matrix ${\bf M}$ is the well known semi-circle law, 
 from which the density of eigenvalues of {\bf C} and {\bf J}=$\mu${\bf
C}$^{-1}$ 
are easily deduced. 
In particular, for the situation we are interested in, one finds
\be
\rho_C(\lambda)=\frac{1}{2\pi\tau}\frac{\sqrt{\tau(4+a)-\lambda}}{\sqrt{
\lambda-a\tau}},
\quad \lambda \in ]a\tau,(4+a)\tau]\label{rho}.
\ee
where $a$ measures the relative amplitude of the diagonal contribution {\bf
D} and $\tau$ the width of the distribution. The distribution of eigenvalues of 
{\bf J}=$\mu${\bf C}$^{-1}$, 
$\rho_J(x)$, can be easily calculated from Eq.(6). Note that in the case where $a=0$, 
$\rho_J(x)$ has a power-law tail 
decaying as $\rho_J(x) \propto x^{-3/2}$. For finite $a$, however,
the maximum eigenvalue of {\bf J} is $\propto 1/a$.

\vskip 0.2cm
\begin{figure}
\centerline{\psfig{figure=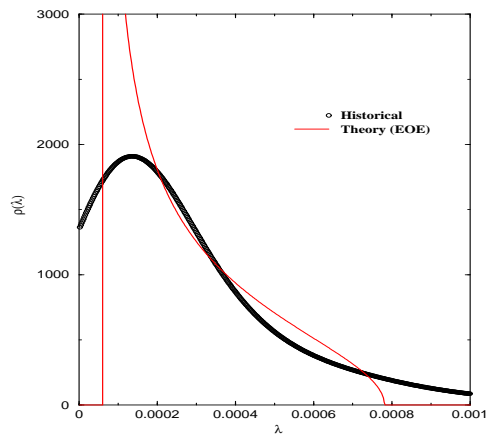,height=5cm,width=5cm}}

\vskip 0.8cm
\caption{Smoothed distribution of eigenvalues of {\bf C}, where the
correlation matrix {\bf C} is extracted from $N=406$ assets of the S\&P500 during 
the years 1991-1996. For comparison we have plotted the best
fit obtained with the theoretical density (6) with $a\simeq .34,
\tau \simeq 1.8\times 10^{-4}$. Results are qualitatively similar in the
case of the London or Zurich Stock Exchanges.} \label{fig1}
\end{figure}

We have studied numerically the density of eigenvalues of the correlation 
matrix of 406 assets of the New-York Stock Exchange ({\sc nyse}), based on daily 
variations during the years 1991-96, for a total of 1500 days. Moreover, by
repeating 
our analysis on different Stock Markets (e.g. London and Zurich) we have
checked
the robustness and universality of our results. The observed density of
eigenvalues is 
shown in Fig.1 together with the best fit obtained by assuming $Q=1$. 
The fitting parameters are $a$ and the width of the semi-circle distribution
$\tau$, 
which are found to be, respectively, $a\simeq .34$ and $\tau \simeq 1.8\times
10^{-4}$.
The deviation from the theoretical curve is partially due to finite $N$
effects
which are expected to smear out the singularities at $\lambda=a\tau$ and
$(4+a)\tau$. It 
might also be that {\bf D} has unequal diagonal elements (this would 
also smear out the singularities of $\rho_J(\lambda)$), or let the number $K$
of 
``explicative factors''  be less than $N$ (a rather likely situation).  Note
that the empirical $\rho_C$ can be rather well fitted by a log-normal distribution,
suggesting that the relevant ensemble has the additional symmetry {\bf C} 
$\leftrightarrow$
{\bf C}$^{-1}$, although we have not been able to justify this property.

The point, however, is not to claim that the density of eigenvalues is
precisely described 
by the above model, but that it represents a reasonable approximation for
our purposes.  
Once the density of eigenstates $\rho_J(\lambda)$ is known, the number of
solutions of the optimization equation, averaged over the matrix ensemble, 
can be computed using tools borrowed from spin glasses and random matrix
theory. As
 discussed in \cite{PaPo}, the main ingredient is the generating function
$G(x)$, defined 
as \cite{IZ}:
\be
\int {\cal D}[\mbox{{\bf J}}]  \exp\left\{\mbox{{ Tr}}\, 
\left(\mbox{{{\bf JA}}}/2\right) \right\} 
\mathop{\simeq}_{N \gg 1} \exp\left\{ N \mbox{{ Tr}}\, 
G\left(\mbox{{{\bf A}}}/N\right)\right\} 
\ee
where ${\bf A}$ is any symmetric $N \times N$ matrix of finite rank, and  ${\cal D}[\mbox{ {\bf
J}}]=
d[\mbox{{\bf J}}] P([\mbox{{\bf J}}])$ is the probability 
measure over the matrix ensemble. The above formula holds for general complex
hermitian 
matrices {\bf J} \cite{IZ}. For example, in the
simplest possible case, where
{\bf J} is extracted from a Gaussian Ensemble, one has
$P([{\bf J}])=\exp\{-\mbox{ Tr}\,( {\bf J}^2/4\sigma^2)\}/Z$ and thus
$G(x)=x^2/4$.
Formally, $G(x)$ can be computed in general from $\rho_J(\lambda)$ using a
series 
of rather involved transformations \cite{Marinari}, which can be somewhat
simplified
by the use of a Hilbert integral transform of the eigenvalue density
\cite{bg97b}. 

In the specific case (\ref{rho}), we have been able to 
compute $G_a(x)$ exactly for $a=0$, with the result  $G_0(x)=-\sqrt{-x}$, and 
perturbatively in the limits $a \ll 1$ and $a \gg 1$. Leaving aside all
mathematical 
details \cite{bg97b}, the results are:
\begin{eqnarray}
G_a(x)&\simeq& \frac{x^2}{4a^4},\qquad a\gg 1, \nonumber \\
G_a(x)&\simeq &-\sqrt{a}g\left(-\frac{x}{a}\right), \qquad a\ll 1,
\end{eqnarray}
where the scaling function $g(u)$ behaves as $g(u)\simeq \sqrt{u}$ for $u \gg
1$ and 
as $g(u) \propto u $ for $u \ll 1$. As a side remark, one can show that for
L\'evy random matrices \cite{Cizeau1} where $\rho(\lambda)$ decays as
$\lambda^{-1-\beta}$ 
for large $\lambda$, the characteristic function behaves as $G(x) \simeq
-(-x)^\beta$ 
for small $x$'s, and for all $\beta < 1$. Consequently,
the problem we are concerned with here ($\beta=1/2$) can be seen as the study
of the 
metastable states in spin glasses with long-range interactions
\cite{Cizeau2,bg97b}. 
 
Once $G_a(x)$ is known, one can write the average number of solutions ${\cal
N}$ of the 
{\sc tap} equations for large $N$ as ${\cal N} \sim \exp \{Nf(a)\}$, where
$f(a)$ is 
obtained from a steepest descent approximation. Indeed, from Eq.(4), one
derives the 
following result, valid for large $N$ \cite{PaPo}:  
\begin{eqnarray}
{\cal N}&=&\left\langle 
\sum_{\{S\}}\prod_{i=1}^N \theta\left(
S_i\sum_{j=1}^N J_{ij}S_j
\right)\right\rangle  \\
&\simeq&\mathop{\max}_{x,y,W,Z}\left\{ \exp\left[
-xZ-yW+G_a(x+\sqrt{y})\right.\right. \nonumber \\
&+& 
G_a(x-\sqrt{y})+\ln\left(\mbox{erfc}\left(-\frac{Z}{2\sqrt{W}} \right)\right)
]\} \\
&\equiv& e^{Nf(a)}.
\end{eqnarray}
where $\theta(u)$ is the usual Heavyside function of its argument. 
From (8), we find that $f(a)$ very rapidly converges to $\ln 2$ (the maximum
allowed number of solutions) when $a$ is large:
\be
f(a)=\ln 2-e^{-a^2/2}\left(\frac{1}{\sqrt{2\pi}a}+o\left(\frac{1}{a^2}\right)
\right),
\ee
a result which we have numerically confirmed by extracting {\bf J} from the
proper 
statistical ensemble. To give an idea, one finds $f(a=1)\simeq .686 \pm .003$,
which is 
already  quite close to $\ln 2=.693147 \ldots$. 

For small $a$, calculations become rather involved, as the 
extremizing set $\{x^*,y^*,W^*,Z^*\}$ corresponds to a minimum and
actually lies on the domain's border.
The final result is that $y^*=x^{*2}$, $W^* = 1/(8 \sqrt{2 x^{*3}})$ and
$Z^* = 1/(4 \sqrt{2 x^{*}})$, which leads to $f(a)\propto a^{1/4}$ for $a \ll 1$. 
We have confirmed this result numerically (see Fig 2). Therefore, for $a=0$,  
$f(a=0)=0$, i.e. the number of metastable 
solutions does not grow exponentially with $N$. This result means that the
presence 
of 
very large eigenvalues prevents 
the existence of many metastable states. This has an interesting 
consequence in the physical context, which was conjectured in \cite{Cizeau2}: 
{\it a spin glass with a broad distribution of couplings cannot sustain many 
ground states}; one can thus expect that the low temperature phase of these
models is rather
different from the `standard picture' \cite{MPV}. We can show that this is
true for 
all $\beta < 1$  \cite{bg97b}; we find in particular that $f(a)\simeq 
K(\beta) a^{\beta/2}$ for small $a$, with a prefactor $K(\beta)$
algebraically  diverging for $\beta \to 1$. For $\beta > 1$, finally, $f(a=0) >0$.
 
Let us now come back to our main theme. In what respect is the above
picture a valid description of the portfolio optimization problem? In the
real 
situation, of course, {\bf C} is given by the ``historical'' correlation
matrix.
If our model is correct, then one should expect the number of optimal
solutions 
to be exponentially large, even for small $a$'s.  By extracting true 
correlation matrices of sizes up to $N=20$ from the available $406$ assets of
the 
{\sc nyse} 
(see above), we have 
indeed numerically found that ${\cal N}_{\sc{NY}}\sim \exp(f_{\sc{NY}}N)$ with $f_{\sc{NY}}=.68\pm .01$. This result does not 
qualitatively change
in the case of the London or Zurich markets.
Despite of the rather low dimensionality of the involved matrices, with 
unavoidable 
finite-size
effects, the above result is quite well reproduced by our theoretical 
model (which is valid at $N\gg 1$). In fact, by using the fitting $(a,\tau)$ 
parameters as 
in Fig.1, we find a theoretical value $f_{\sc{TH}}= .63 \pm 0.02$, not too far
from the 
real one. Similarly by extracting {\bf C} from 172 stocks of the London Stock
Exchange ({\sc LSE}) we have found an empirical $f_{\sc{LSE}}=.43 \pm .09$ and a 
corresponding theoretical value $f_{\sc{TH}}=.32 \pm .07$. 
Remarkably, our admittedly approximate description of the density of
states of the matrix {\bf J} seems to account satisfactorily for the observed 
number of ``optimal portfolios''. 

\vskip 0.2cm                                 
\begin{figure}
\centerline{\psfig{figure=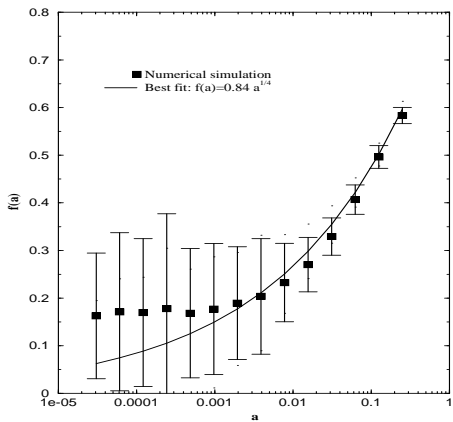,height=5cm,width=5cm}}

\vskip 0.8cm
\caption{The theoretical exponent $f(a)$ plotted vs $a$ ($a \ll 1$) when
{\bf C} is extracted from the {\sc EOE}, using an exact 
enumeration procedure to find the numbers of  solutions up to $N=27$. 
Finite size effects are clearly present for the smallest
values of $a$.  The theoretical line $K a^{1/4}$ is plotted for comparison, with $K \sim .84$.} \label{fig2}
\end{figure}

By means of a combination  of analytical and numerical arguments, we have 
thus shown that the number of optimal portfolios in  futures markets
(where the constraint on the weights is non linear) grows exponentially 
with the number of assets. On the example of U.S. stocks, we find that an 
optimal portfolio with $100$ assets can be composed in $\sim 10^{29}$ 
different ways ! Of course, the above calculation counts all local optima, 
disregarding the
associated value of the residual risk ${\cal R}$. One could extend the
above calculations to obtain the number of solutions for a given ${\cal R}$.
Again, in analogy with \cite{PaPo}, we expect this number to grow as
$\exp Nf({\cal R},a)$, where $f({\cal R},a)$ has a certain parabolic 
shape, which goes to zero for a certain `minimal' risk ${\cal R}^*$. But for
any small interval around ${\cal R}^*$, there will already be an exponentially
large number of solutions. The most interesting feature, with particular
reference to applications in economy, finance or social sciences, is that
all these quasi-degenerate solutions can be very far from each other: in 
other words, it can be rational to follow two totally different strategies! 
Furthermore, these solutions are usually `chaotic' in the sense that a small
change of the matrix {\bf J}, or the addition of an extra asset, completely 
shuffles the order of the solutions (in terms of their risk). Some solutions 
might even disappear, while other appear. 
It is worth pointing out that the above scenario is not restricted 
to portfolio theory, but is germane to a variety of situations. 
For example, it is well known in Game Theory that
each player has a different utility function he must maximize in order
to get the best profit. When, in addition, nonlinear contraints are present, 
we expect a similar proliferation of solutions.  
As emphasized above, the existence of an exponentially large
number of solutions forces one to rethink the very concept of 
rational decision making.

Aknowledgements: We want to thank J.P. Aguilar, R. Cont and L. Laloux for discussions. The
{\sc lse} and Zurich data were obtained from the Financial Times and the {\sc nyse} from S\&P Compustat.

\end{document}